\documentclass[%
 reprint,
superscriptaddress,
 amsmath,amssymb,
 aps,
prl,
]{revtex4-2}

\usepackage{graphicx}
\usepackage{dcolumn}
\usepackage{bm}

\begin{document}


\title{Diffusion Monte Carlo calculations of fully-heavy compact hexaquarks}

\author{J. M. Alcaraz-Pelegrina}
\email{fa1alpej@uco.es}
\affiliation{Departamento de Sistemas F\'{\i}sicos, Qu\'{\i}micos
y Naturales, Universidad Pablo de
Olavide, Carretera de Utrera km 1, E-41013 Sevilla, Spain}
\affiliation{Departamento de F\'{\i}sica.
Universidad de C\'ordoba, Campus de Rabanales, Edif. C2 
E-14071 C\'ordoba, Spain}

\author{M. C. Gordillo}
\email{cgorbar@upo.es}
\affiliation{Departamento de Sistemas F\'{\i}sicos, Qu\'{\i}micos
y Naturales, Universidad Pablo de
Olavide, Carretera de Utrera km 1, E-41013 Sevilla, Spain}
\affiliation{Instituto Carlos I de F\'{\i}sica Te\' orica y Computacional, Universidad de Granada, E-18071 Granada, Spain.}

\date{\today}

\begin{abstract}
We used a diffusion Monte Carlo technique to describe the properties of fully-heavy compact arrangements (no dibaryon molecules) including six quarks and no 
antiquarks within the framework of a constituent quark model. 
Only arrangements whose wavefunctions were eigenvectors of $L^2$ with eigenvalue $\ell$ = 0 were taken into account, what means that we only considered the subset of all the possible 
color-spin combinations that make   
the total wavefunctions antisymmetric with respect to the interchange
of any two quarks of the same type. 
In all cases, the masses of the six-quark arrangements are larger that the ones corresponding to the sum of any of the two baryons 
we can split them into, but smaller that the ones for a set of six isolated quarks, i.e., all them are bound systems. The analysis of their structure indicates that 
all the hexaquarks considered in this work are compact objects, except the $cccbbb$, that appears to be a loose association of two baryons for all the possible spin values.                    
\end{abstract}

\maketitle


\section{Introduction}
                 

Even tough any six-quark combination can be called an hexaquark, one can think of at least two types of arrangements that can bear that name: a compact six-quark
cluster and a loosely bound dibaryon \cite{reviewvalcarce,clement,clement2}. The only dibaryons experimentally produced so far are the deuteron \cite{1932}, together with 
the well-established $d^*(2380)$ resonance \cite{d2380b,d2380c,d2380d,d2380e,d2380}. While the first one appears to be an association of two baryons,  
there is no consensus about the structure of the second \cite{clement2}. In any case,  
the majority of the searches and the theoretical investigations were made in the flavored or strange sectors \cite{clement}. However, there are 
some recent works that tackle the problem of partially heavy (see Ref.  \onlinecite{prc2022} and references therein) and fully heavy \cite{wang,prlbaryons,suma,vijande,hexa22,hexa22b} dibaryons. This last case is specially challenging since there is no experimental data on fully heavy baryons, even 
though the X(6900) state has been proposed to be an ensemble of two $c$ and two $\bar c$ quarks \cite{x6900}.  

What all the previous theoretical investigations about hexaquarks have in common is that almost every one of them deals with an association of two baryons or three diquarks.    
In this work, we transcend that approximation and deal with color-spin functions that are not subject to that limitation, using the constituent
quark model to describe compact structures made up of six heavy quarks.      
The underlying assumption behind that model 
is that any arrangement of quarks can be described by a Hamiltonian of the 
type \cite{reviewvalcarce}: 
\begin{equation}
H = \sum_{i=1}^{N_q} \left( m_i+\frac{\vec{p}^{\,2}_i}{2m_i}\right) + \sum_{i<j}^{N_q} V(r_{ij}) \,,
\label{eq:Hamiltonian}
\end{equation}
where $N_q$ is the number of quarks, 
that in this work will be fixed to six, 
and $m_{i}$ and $\vec{p}_i$ are the mass and momentum of the $i$ quark.
The two-body potential, $V(r_{ij})$, depends only on the distance between quarks, $r_{ij}$, and can be written as the sum of one-gluon exchange  
and confinement contributions. The first of 
those can expressed as:
\begin{equation}
V_{\text{OGE}}(r_{ij}) = \frac{1}{4} \alpha_{s} (\vec{\lambda}_{i}\cdot
\vec{\lambda}_{j}) [\frac{1}{r_{ij}} - \frac{2\pi}{3m_{i}m_{j}} \delta^{(3)}(r_{ij}) (\vec{\sigma}_{i}\cdot\vec{\sigma}_{j}) ] \,,
\end{equation}
and includes both Coulomb and hyperfine terms.  Here, $\vec{\lambda}$ and $\vec{\sigma}$ are the  Gell-Mann and Pauli matrices, respectively, 
and account for the color and spin degrees of freedom of the constituent particles.   
The notorious difficulty of dealing numerically with the Dirac delta function was overcome in the standard 
fashion, i.e., by replacing it by a smeared out function \cite{Semay:1994ht, SilvestreBrac:1996bg,prdyo1}.  
The contribution of multi-gluon exchanges is introduced effectively by a linear confining potential 
proportional to the distance between quarks~\cite{reviewvalcarce}. This means:  
\begin{equation}
V_{\text{CON}}(\vec{r}_{ij}) = (b\, r_{ij} + \Delta) (\vec{\lambda}_{i}\cdot \vec{\lambda}_{j}). 
\end{equation}
In principle, this description can be applied to any ensemble of quarks and/or antiquarks, even tough given the non-relativistic nature of the 
Hamiltonian in Eq. \ref{eq:Hamiltonian}, one expects it to afford a more reasonable description when the cluster contains one or more heavy ($c$ and $b$)
quarks.  

Even though the form of the potential terms is more or less standard within the framework of the quark model, the parameters that define them can vary. 
In this work, and to be coherent in our comparisons with previous results in heavy baryons \cite{prdyo1}, we used 
the so-called AL1 potential proposed by Silvestre-Brac and Semay in Refs.~\cite{Semay:1994ht, SilvestreBrac:1996bg}, works from which all the 
necessary parameters were obtained. The properties of the hadrons computed with this potential were found to be in good agreement with 
experimental data, when available \cite{prdyo1}. 

\section{Method}

Once defined the Hamiltonian that describes the system, we have to solve the corresponding Schrodinger equation in order to obtain the properties of
the hexaquarks.  To do so, we chose a diffusion Monte Carlo (DMC) algorithm \cite{kalos,boro94,hammond,spin-orbita}.  This is an stochastic technique that allows us to 
obtain an upper bound for the energies of the ground state of an arrangement of fermions (as quarks are) within the statistical uncertainties of any Monte Carlo scheme. 
The only drawback of the method is that  an initial approximation to the real many-body wavefunction of the set of quarks, the so-called {\em trial function} has to be provided.  
This function has to contain all the information known a priori about the system.  Since every quark has, besides its position, associated a value of spin and color,  we chose the simple expression
\cite{prdyo1}:
\begin{eqnarray}
\Psi({\bf r_1, r_2},   \ldots,  {\bf  r_6}, s_1,s_2,  \ldots, s_6,c_1,c_2,\ldots, c_6) = \nonumber \\
\Phi ({\bf r_1, r_2},   \ldots,  {\bf  r_6}) \nonumber \\ 
\left[\chi_s (s_1,s_2,  \ldots, s_6) \bigotimes \chi_c (c_1,c_2,\ldots, c_6) \right] ,
\end{eqnarray}
where ${\bf r_i}$, $s_i$ and $c_i$ stand for the position, spin and color or quark $i$.  We defined $\Phi$ as: 
\begin{equation} \label{radial}
\Phi ({\bf r_1, r_2},   \ldots,  {\bf  r_6}) = \prod_{i=1}^{N_q} \exp(-a_{ij} r_{ij}),
\end{equation} 
i.e.,  as a product of the ground state solutions of Schr\"odinger equations including only a Coulombic term for as many independent pair of quarks as we have in the hexaquark.  
In that spirit, $a_{ij}$ is chosen to take care of the cusp conditions, i.e., to avoid the divergence of the derivatives of the trial
function when $r_{ij} \rightarrow$ 0.  $\Phi$ is also an eigenvalue of the total angular momentum of the hexaquark, $L^2$, with eigenvalue $\ell$ = 0.  
No other alternatives to the form of the radial part of the trial function were considered in this work.   $\chi_s$ and $\chi_c$ are linear combinations of functions including a value for the spin and color for every quark.  This can be done in a step-by-step procedure similar to the used in Ref. \onlinecite{hexa22}, by using the Clebsch-Gordan coefficients of the corresponding color and spin groups.  However,  this is extremely cumbersome, and eventually it will become impossible to apply for progressively larger ensembles of quarks.  In this work, we propose an alternative via that can be easily automated and is, in principle, scalable for any size of the system.   Obviously, it is also completely equivalent to the standard approach.

First, we started by calculating the eigenvectors of the color and spin
operators, defined as: 
\begin{equation}
F^2 = \left(\sum_{i=1}^{N_q} \frac{\lambda_i}{2} \right)^2  
\end{equation}  
and 
\begin{equation}
S^2 = \left(\sum_{i=1}^{N_q} \frac{\sigma_i}{2}\right)^2.
\end{equation}
The color space spans all possible color combinations from $|rrrrrr\rangle$ to
$|gggggg\rangle$, while each spin vector is made up of all the possible sets of six spin values.   Obviously, in the case of $F^2$,  we keep only the five wavefunctions with eigenvalue equal to zero \cite{fivefunctions,5f2,5f3,5f0}, i.e., the one that are $colorless$.  The number of spin functions are 5 for $S=0,$  27 for $S=1$,  25 for $S=2$ and 7 for $S=3$. Once we have those eigenvectors, we can construct the color-spin functions as $\chi_s \bigotimes \chi_c$ products.  However, to describe adequately a system of quarks, and 
given that Eq. \ref{radial} is symmetric with respect to the exchange of any two quarks, the necessary antisymmetry of the total wavefunction has to be included in the $\chi_s \bigotimes \chi_c$ product., i.e., we have to produce a linear combination of the spin-color functions antisymmetric with respect to the interchange of any identical quarks.  
To do so, we apply the antisymmetry operator:
\begin{equation}
\mathcal{A} = \frac{1}{N} \sum_{{\alpha}=1}^N (-1)^P \mathcal{P_{\alpha}}
\label{anti}
\end{equation}
to that color-spin set of functions.  Here, $N$ is the number of possible permutations of the quarks indexes, $P$ is the order of the permutation, 
and $\mathcal{P_{\alpha}}$ represents the matrices that define those permutations. 
For instance, if we have six identical quarks, $N =$ 6! = 720.   
Once constructed the matrix derived from the operator in Eq. \ref{anti},  we have to check if we can find any eigenvector with eigenvalue equal to one.  If this is not possible, then 
a six-quark arrangement with a radial part given by Eq. \ref{radial} does not exist.  On the other hand,  when one or several of those functions fulfill the antisymmetry requirements,  we use those combinatinos as input in the DMC algorithm in the way described in 
Ref. \onlinecite{prdyo1}.

\section{Results}

Once we have defined the radial and color-spin functions corresponding to each of the systems we are interested in, we can apply the DMC technique to obtain their masses in the same 
way described in previous literature \cite{prdyo1}. This is basically a recipe that uses a combination of a standard DMC method and weights the results by a Green-function projection 
that depends on the color-spin part of the Hamiltonian \cite{spin-orbita,prdyo1}. However, a single weight algorithm based on the Green-function Monte Carlo for nuclei described 
in Ref. \onlinecite{carlson} can also be used, providing exactly the same 
results.  This DMC method allows us to calculate not only the masses but the structure of the 
hexaquarks via the radial distribution function, $\rho(r_{ij})$, that gives us an idea of the how probable is for a pair of quarks to be at a given distance, $r_{ij}$, from each other
\cite{prdyo1,prdyo2}. This is possible since the DMC calculations take into account in full the correlations between particles.

\begin{table} 
\caption{\label{tab:table1} Masses of the hexaquarks considered in this work in MeV. The error bars are $\pm$ 2 MeV in all cases. Also included are the masses of all the possible combination of 
baryons the hexaquarks can split into.  Some of those masses were  taken from Ref. \onlinecite{prdyo1} while the remaining ones were calculated in this work 
}
\begin{tabular}{ccccc} \hline 
Hexaquark &  $0^+$  & $1^+$ & $2^+$ & $3^+$    \\ \hline 
$cccccc$ & 9904 &  --   & -- &--  \\                                   
$bbbbbb$ & 29114 &  --   & -- & --  \\                                   
$cccccb$ & 13141 & 13122   & -- & -- \\
$bbbbbc$ & 25955 & 25913   & -- & -- \\
$ccccbb$ & 16280 & 16296 & 16279 & -- \\
$bbbbcc$ & 22689 & 22703 & 22683 & --\\
$cccbbb$ & 19216 & 19221 & 19197 & 19193\\ \hline
Dibaryons &  $1/2^+ + 1/2^+$ & $3/2^+ + 1/2^+$  &  $3/2^+ + 3/2^+$ \\ \hline 
$ccc$ + $ccc$  & -- & -- & 9596$^*$ \\
$bbb$ + $bbb$  & -- & -- & 28796$^*$ \\
$ccc$ + $ccb$  & -- & 12816$^*$ & 12844 \\
$bbb$ + $cbb$  & -- & 25613$^*$ & 25645 \\
$ccc$ + $bbb$  & -- & -- & 19197$^*$ \\
$ccc$ + $cbb$  & -- & 16013$^*$ & 16045\\ 
$ccb$ + $ccb$  &  16036$^*$  & 16064  &  16092 \\
$bbb$ + $ccb$  &    -- & 22416$^*$ & 22444 \\  
$cbb$ + $cbb$  &  22430$^*$ & 22462 & 22494 \\ \hline 
$^*$Ref. \onlinecite{prdyo1}
\end{tabular} 
\label{table1}
\end{table}

The masses of all the possible all-heavy hexaquarks made up of six quarks and no antiquarks obtained by DMC are given in Table \ref{table1}. We include also the masses of all the possible pairs 
of baryons compatible with the composition of the hexaquarks. A couple of things are immediately apparent. First, and given 
that the bare masses of the $c$ and $b$ quarks are 1836 and 5227 MeV \cite{SilvestreBrac:1996bg,prdyo1}, respectively, all the hexaquarks have a smaller mass  than those  
of their constituents, i.e., all the hexaquarks are bound systems. However, a glimpse at the second part of Table \ref{table1} also indicates that their masses are also larger
than the ones for any couple of baryons they can be divided into. This means that any hexaquark is unstable with respect to its splitting into two baryons.  

\begin{figure}
\begin{center}
\includegraphics[width=0.8\linewidth]{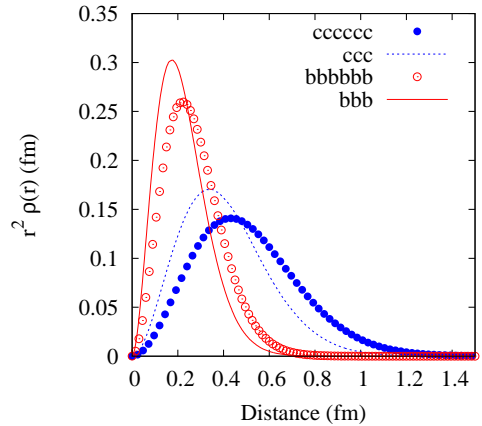}
\caption{Radial distribution functions for the $cc$ and $bb$ quark pairs in the $cccccc$ and $bbbbbb$ hexaquarks for $S$=0 (symbols). Also shown for comparison the same functions for
the $ccc$ and $bbb$ baryons in their ground states taken from Ref. \onlinecite{prdyo1} (lines).   
}
\label{fig1}
\end{center}
\end{figure}

\begin{figure}
\begin{center}
\includegraphics[width=0.8\linewidth]{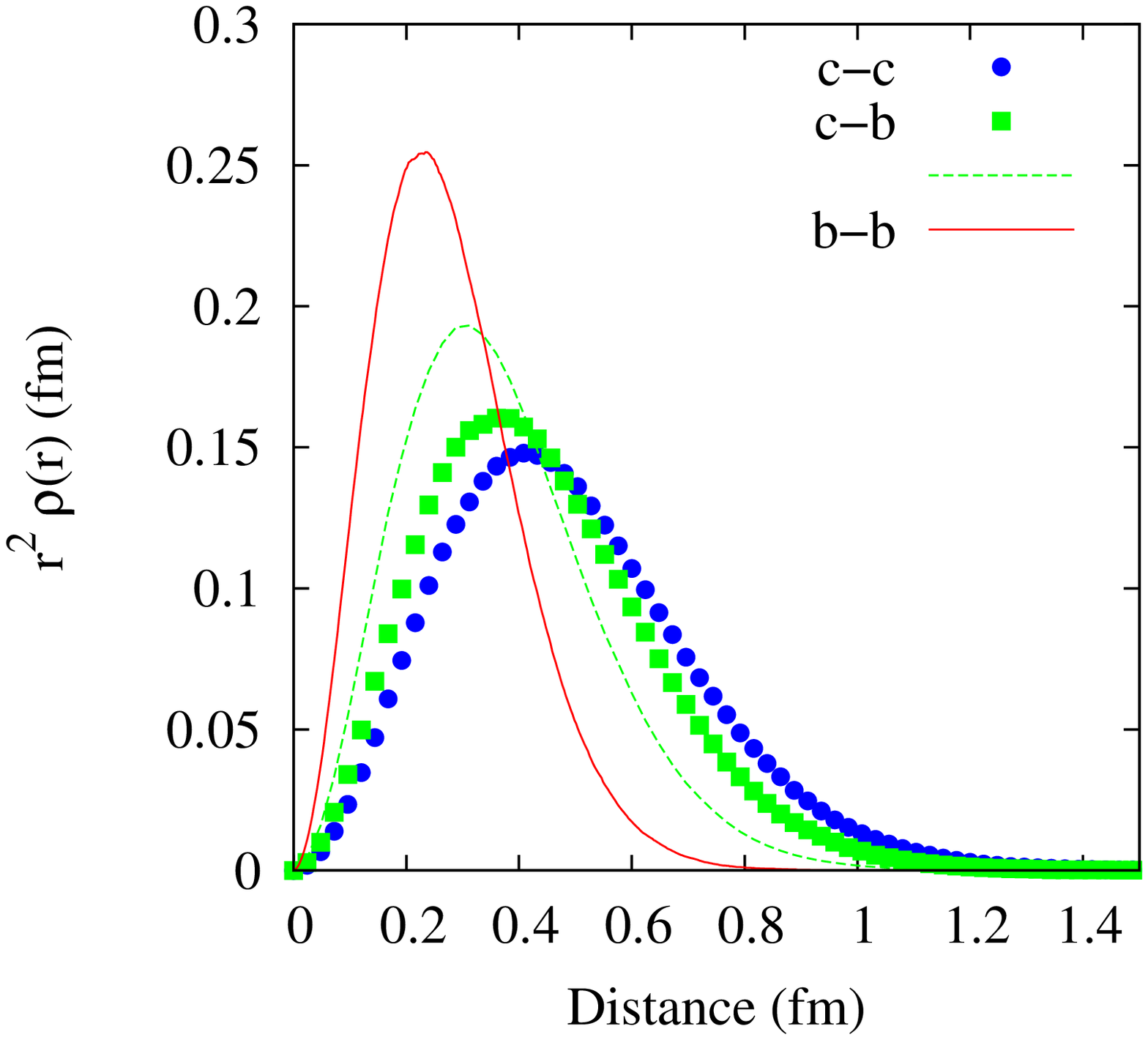}
\caption{Radial distribution functions for the $cc$, $cb$ and $bb$ quark pairs in the $cccccb$ (symbols) and $bbbbbc$ (lines) hexaquarks. 
}
\label{fig2}
\end{center}
\end{figure}

\begin{figure}
\begin{center}
\includegraphics[width=0.8\linewidth]{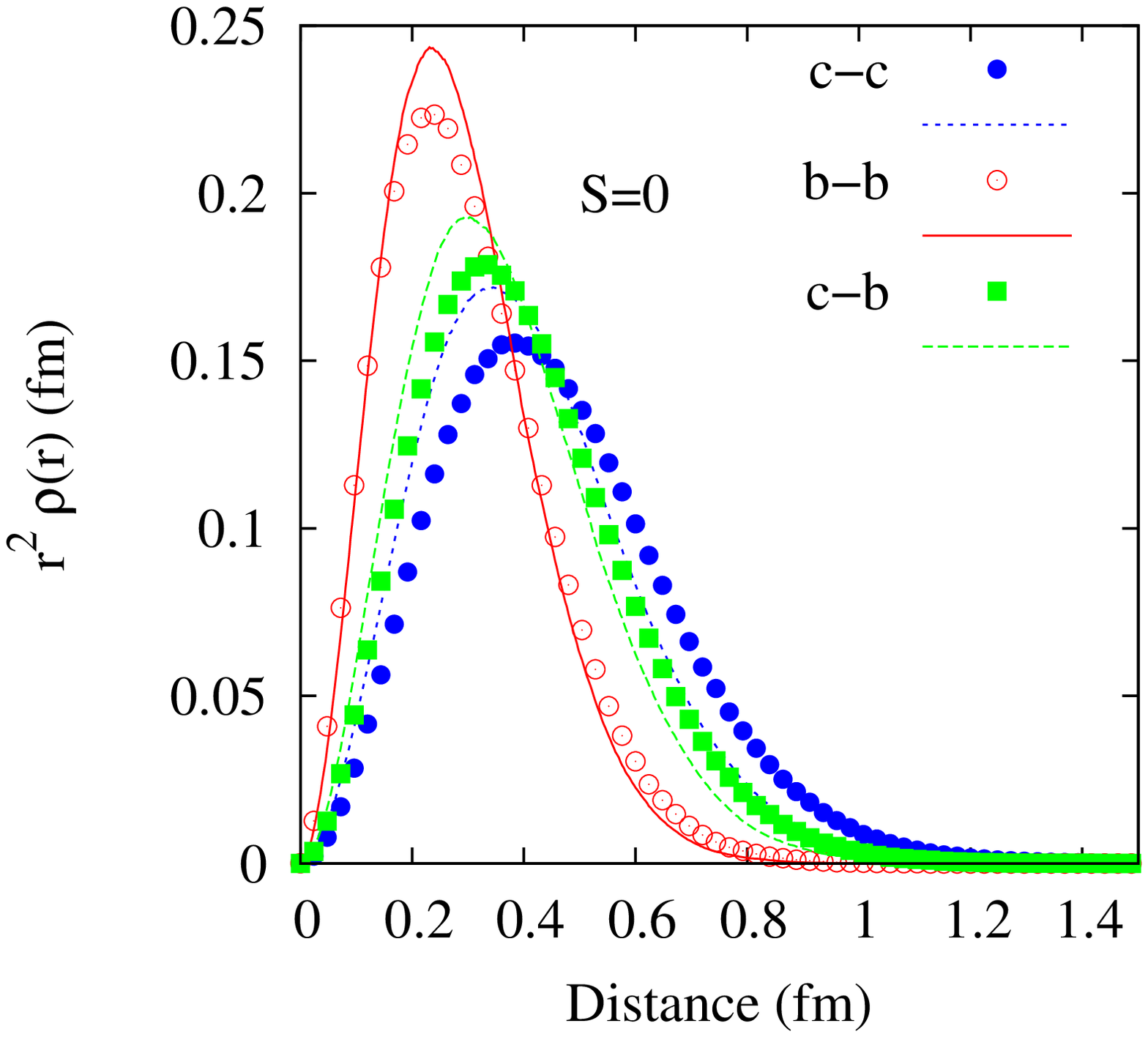}
\caption{Radial distribution functions for the $cc$, $cb$ and $bb$ quark pairs in the $ccccbb$ (symbols) and $bbbbcc$ (lines) hexaquarks for $S$=0. 
}
\label{fig3}
\end{center}
\end{figure}

\begin{figure}
\begin{center}
\includegraphics[width=0.8\linewidth]{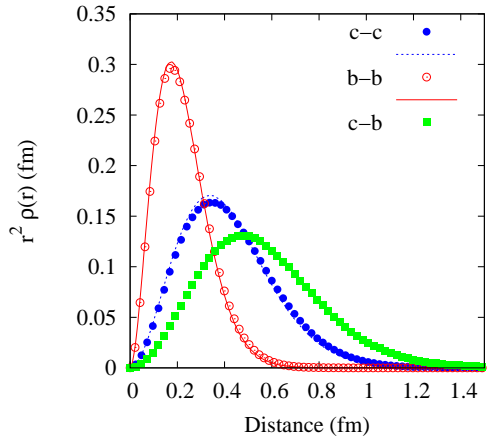}
\caption{Radial distribution functions for the quark pairs inside the $cccbbb$ $S=0$ hexaquark (symbols) and the $ccc$ and $bbb$ baryons (lines).   
}
\label{fig4}
\end{center}
\end{figure}

As indicated above, the DMC technique is able to give us an idea about the distribution of the particles inside the different hexaquarks via the radial distribution functions. Those 
will be displayed in Figs. \ref{fig1}, \ref{fig2}, \ref{fig3} and \ref{fig4}. Fig. \ref{fig1} represents the case in which all the quarks are of the same type. This means that we
only can have pair distributions corresponding to $cc$ and $bb$ pairs for the $cccccc$ and $bbbbbb$ systems, respectively.  In principle, we have fifteen of such pairs, and the 
results represented are normalized averages of those fifteen functions. In that figure we also show, for comparison, the same radial distributions but for the corresponding baryons, 
taken from Ref. \onlinecite{prdyo1}. What we can see is that we have compact objects with a single maximum in the $cc$ o $bb$ squared distance, maximum that is larger than in the 
case of the corresponding baryons. We see also that the position of those maxima depends on the mass of the quarks involved, being larger for the least massive ($cc$) pair.            

Compact objects with a single maximum in the $cc$, $cb$ and $bb$ radial distribution functions are also the $cccccb$, and $bbbbbc$ hexaquarks, whose radial distribution functions 
are given in Fig. \ref{fig2}. 
In addition, and as in the previous arrangements, we see that the maxima in the $r^2 \rho(r)$ distributions is inversely proportional to the pair involved. This means smaller for the $bb$ 
distribution and larger for the $cc$ pair, with the $cb$ case in-between. This fact can be seen also for the $ccccbb$ and $bbbbcc$ six-quark clusters, shown in Fig. \ref{fig3}. 
There, we display only the results for $S=0$, since the remaining cases are qualitatively similar. The only difference between them is their relative sizes, that can be measured
via their mean squared radii \cite{SilvestreBrac:1996bg,prdyo1}. For the $ccccbb$ clusters, those radii are 0.073 ($S$ = 0) and 0.081 fm$^2$ ($S$= 1 and $S$=2) with an error bar of $\pm$ 
0.002 fm$^2$ in all cases. The corresponding values for the $bbbbcc$ hexaquark are 0.050 ($S$= 0) and 0.054 fm$^2$ ($S$= 1 and $S$=2), with the same  error bars as in the previous case.    
For the sake of comparison, the mean squared radii for the $cccccc$ and $bbbbbb$ arrangements  are 0.131 and 0.037 $\pm$ 0.002 fm$^2$, respectively, and for the $cccccb$ and $bbbbbc$
sets, 0.101 and 0.044 $\pm$ 0.002 fm$^2$ for $S=0$ and 0.104 and 0.045 $\pm$ 0.002 fm$^2$ for $S=1$ . Obviously, the larger the number of $c$ quarks involved, the bigger the hexaquark.     

The remaining hexaquark, $cccbbb$ is qualitatively different from all of the other ones. This can be seen in its structure, displayed in Fig. \ref{fig4} for $S=0$. The other cases as qualitatively similar and not displayed by simplicity.  There, we represent the 
radial distribution functions for the hexaquark (symbols) together with the ones corresponding to the $ccc$ and $bbb$ bayrons. Two things are immediately apparent: even though the
total spread of the functions is similar to those of the previously shown clusters, the position of the maxima is changed with respect to them. Instead of having $cc > cb > bb$, 
the order is $bb > cc > cb$. In addition, the $cc$ and $bb$ distributions are nearly identical to those corresponding to the baryons displayed (lines). This 
strongly suggests that we have two independent baryons close to each other and not a compact hexaquark. This interpretation is supported by the fact that in none of the other hexaquaks 
the mass is so close to those of the baryons they can be splitted into, detailed in Table \ref{table1}.     

\section{Conclusions}

In this work we have described all six heavy quark ensembles using trial functions whose only constraint was their antisymmetry with respect to the exchange of any two identical 
fermions. In particular, no grouping in sets of baryons or diquarks was considered.  The antisymmetry in the color-spin part of the wavefunction was introduced by using a direct diagonalization of the antisymmetry operator, instead of building up the functions via Clebsh-Gordan coefficients. In addition, the use 
of a DMC technique made easier to tackle the full six-particle problem since, 
in contraposition to other algorithms such as the Gaussian expansion 
method \cite{hiyama}, DMC was designed to deal with many-body problems \cite{kalos}.      
Not only that, but DMC gives us information about the structure of the hexaquarks via the radial distribution functions and mean square radii.  
Those radial distribution functions, depicted in Figs. \ref{fig1} to \ref{fig4}, allows us to say that all the hexaquarks considered here are compact structures, except the 
$cccbbb$ one, that appears to be a juxtaposition of $ccc$ + $bbb$. However, the values of the masses shown in Table \ref{table1}, also indicate that the compact bags are 
metastable with respect to their splitting in any of the two baryons compatible with their composition. Those mass differences are in the range 200-300 MeV in all cases except
for the $cccbbb$ hexaquark, very close to the  $ccc$+$bbb$ value. This suggests that that hexaquark does not exist as a compact structure, not even a metastable one,
since even the a priori compact structure derived from the eigenvalues of $F^2$ and $S^2$ is split into two well-defined baryons. This is the same conclusion obtained in 
Ref. \onlinecite{vijande}.

\begin{acknowledgments}

We acknowledge financial support from Ministerio de
Ciencia e Innovación MCIN/AEI/10.13039/501100011033 
(Spain) under Grant No. PID2020-113565GB-C22
and from Junta de Andaluc\'{\i}a group PAIDI-205.  J.M.A.P. acknowledges by a fellowship from the Spanish Ministerio de Universidades under a grant from \emph{Plan de Recuperación, Transformación y Resiliencia} funded by the European Union-NextGenerationEU.
We also acknowledge the use of the C3UPO computer facilities at the Universidad
Pablo de Olavide.
\end{acknowledgments}

\bibliography{h3}

\begin{thebibliography}{31}%
\makeatletter
\providecommand \@ifxundefined [1]{%
 \@ifx{#1\undefined}
}%
\providecommand \@ifnum [1]{%
 \ifnum #1\expandafter \@firstoftwo
 \else \expandafter \@secondoftwo
 \fi
}%
\providecommand \@ifx [1]{%
 \ifx #1\expandafter \@firstoftwo
 \else \expandafter \@secondoftwo
 \fi
}%
\providecommand \natexlab [1]{#1}%
\providecommand \enquote  [1]{``#1''}%
\providecommand \bibnamefont  [1]{#1}%
\providecommand \bibfnamefont [1]{#1}%
\providecommand \citenamefont [1]{#1}%
\providecommand \href@noop [0]{\@secondoftwo}%
\providecommand \href [0]{\begingroup \@sanitize@url \@href}%
\providecommand \@href[1]{\@@startlink{#1}\@@href}%
\providecommand \@@href[1]{\endgroup#1\@@endlink}%
\providecommand \@sanitize@url [0]{\catcode `\\12\catcode `\$12\catcode
  `\&12\catcode `\#12\catcode `\^12\catcode `\_12\catcode `\%12\relax}%
\providecommand \@@startlink[1]{}%
\providecommand \@@endlink[0]{}%
\providecommand \url  [0]{\begingroup\@sanitize@url \@url }%
\providecommand \@url [1]{\endgroup\@href {#1}{\urlprefix }}%
\providecommand \urlprefix  [0]{URL }%
\providecommand \Eprint [0]{\href }%
\providecommand \doibase [0]{https://doi.org/}%
\providecommand \selectlanguage [0]{\@gobble}%
\providecommand \bibinfo  [0]{\@secondoftwo}%
\providecommand \bibfield  [0]{\@secondoftwo}%
\providecommand \translation [1]{[#1]}%
\providecommand \BibitemOpen [0]{}%
\providecommand \bibitemStop [0]{}%
\providecommand \bibitemNoStop [0]{.\EOS\space}%
\providecommand \EOS [0]{\spacefactor3000\relax}%
\providecommand \BibitemShut  [1]{\csname bibitem#1\endcsname}%
\let\auto@bib@innerbib\@empty
\bibitem [{\citenamefont {Valcarce}\ \emph {et~al.}(2005)\citenamefont
  {Valcarce}, \citenamefont {{\relax H. Garcilazo, F Fernández}},\ and\
  \citenamefont {González}}]{reviewvalcarce}%
  \BibitemOpen
  \bibfield  {author} {\bibinfo {author} {\bibfnamefont {A.}~\bibnamefont
  {Valcarce}}, \bibinfo {author} {\bibnamefont {{\relax H. Garcilazo, F
  Fernández}}},\ and\ \bibinfo {author} {\bibfnamefont {P.}~\bibnamefont
  {González}},\ }\bibfield  {title} {\bibinfo {title} {{Quark model study of
  few-baryon systems}},\ }\href@noop {} {\bibfield  {journal} {\bibinfo
  {journal} {Rep. Prog. Phys.}\ }\textbf {\bibinfo {volume} {68}},\ \bibinfo
  {pages} {965} (\bibinfo {year} {2005})}\BibitemShut {NoStop}%
\bibitem [{\citenamefont {Clement}(2017)}]{clement}%
  \BibitemOpen
  \bibfield  {author} {\bibinfo {author} {\bibfnamefont {H.}~\bibnamefont
  {Clement}},\ }\bibfield  {title} {\bibinfo {title} {{On the History of
  Dibaryons and their Final Observation}},\ }\href
  {https://doi.org/10.1016/j.ppnp.2016.12.004} {\bibfield  {journal} {\bibinfo
  {journal} {Prog. Part. Nucl. Phys.}\ }\textbf {\bibinfo {volume} {93}},\
  \bibinfo {pages} {195} (\bibinfo {year} {2017})}\BibitemShut {NoStop}%
\bibitem [{\citenamefont {Clement}\ and\ \citenamefont
  {Skorodko}(2021)}]{clement2}%
  \BibitemOpen
  \bibfield  {author} {\bibinfo {author} {\bibfnamefont {H.}~\bibnamefont
  {Clement}}\ and\ \bibinfo {author} {\bibfnamefont {T.}~\bibnamefont
  {Skorodko}},\ }\bibfield  {title} {\bibinfo {title} {{Dibaryons: Molecular
  versus compact hexaquarks}},\ }\href@noop {} {\bibfield  {journal} {\bibinfo
  {journal} {Chinese Phys. C}\ }\textbf {\bibinfo {volume} {45}},\ \bibinfo
  {pages} {022001} (\bibinfo {year} {2021})}\BibitemShut {NoStop}%
\bibitem [{\citenamefont {Urey}\ \emph {et~al.}(1932)\citenamefont {Urey},
  \citenamefont {{\relax F. G. Brickwedde}},\ and\ \citenamefont
  {Murphy}}]{1932}%
  \BibitemOpen
  \bibfield  {author} {\bibinfo {author} {\bibfnamefont {H.~C.}\ \bibnamefont
  {Urey}}, \bibinfo {author} {\bibnamefont {{\relax F. G. Brickwedde}}},\ and\
  \bibinfo {author} {\bibfnamefont {G.~M.}\ \bibnamefont {Murphy}},\ }\bibfield
   {title} {\bibinfo {title} {{A Hydrogen Isotope of Mass 2}},\ }\href
  {https://doi.org/10.1103/PhysRev.39.164} {\bibfield  {journal} {\bibinfo
  {journal} {Phys. Rev.}\ }\textbf {\bibinfo {volume} {39}},\ \bibinfo {pages}
  {164} (\bibinfo {year} {1932})}\BibitemShut {NoStop}%
\bibitem [{\citenamefont {et~al. (CELSIUS-WASA~Collaboration}(2009)}]{d2380b}%
  \BibitemOpen
  \bibfield  {author} {\bibinfo {author} {\bibfnamefont {M.~B.}\ \bibnamefont
  {et~al. (CELSIUS-WASA~Collaboration}},\ }\bibfield  {title} {\bibinfo {title}
  {{Double-Pionic Fusion of Nuclear Systems and the ‘‘ABC’’
  Effect:Approaching a Puzzle by Exclusive and Kinematically Complete
  Measurements}},\ }\href {https://doi.org/10.1103/PhysRevLett.102.052301}
  {\bibfield  {journal} {\bibinfo  {journal} {Phys. Rev. Lett.}\ }\textbf
  {\bibinfo {volume} {102}},\ \bibinfo {pages} {052301} (\bibinfo {year}
  {2009})}\BibitemShut {NoStop}%
\bibitem [{\citenamefont {et~al. (WASA-at COSY~Collaboration}(2011)}]{d2380c}%
  \BibitemOpen
  \bibfield  {author} {\bibinfo {author} {\bibfnamefont {P.~A.}\ \bibnamefont
  {et~al. (WASA-at COSY~Collaboration}},\ }\bibfield  {title} {\bibinfo {title}
  {{Abashian-Booth-Crowe Effect in Basic Double-Pionic Fusion: A New
  Resonance?}},\ }\href {https://doi.org/10.1103/PhysRevLett.106.242302}
  {\bibfield  {journal} {\bibinfo  {journal} {Phys. Rev. Lett.}\ }\textbf
  {\bibinfo {volume} {106}},\ \bibinfo {pages} {242302} (\bibinfo {year}
  {2011})}\BibitemShut {NoStop}%
\bibitem [{\citenamefont {et~al. (WASA-at
  COSY~Collaboration}(2013{\natexlab{a}})}]{d2380d}%
  \BibitemOpen
  \bibfield  {author} {\bibinfo {author} {\bibfnamefont {P.~A.}\ \bibnamefont
  {et~al. (WASA-at COSY~Collaboration}},\ }\bibfield  {title} {\bibinfo {title}
  {{Isospin decomposition of the basic double-pionic fusion in the region of
  the ABC effect}},\ }\href {https://doi.org/10.1016/j.physletb.2013.03.019}
  {\bibfield  {journal} {\bibinfo  {journal} {Phys. Lett. B}\ }\textbf
  {\bibinfo {volume} {721}},\ \bibinfo {pages} {229} (\bibinfo {year}
  {2013}{\natexlab{a}})}\BibitemShut {NoStop}%
\bibitem [{\citenamefont {et~al. (WASA-at
  COSY~Collaboration}(2013{\natexlab{b}})}]{d2380e}%
  \BibitemOpen
  \bibfield  {author} {\bibinfo {author} {\bibfnamefont {P.~A.}\ \bibnamefont
  {et~al. (WASA-at COSY~Collaboration}},\ }\bibfield  {title} {\bibinfo {title}
  {{Measurement of the pn -> pp $\pi^0 \pi^-$ reaction in isearch for the
  recently observed resonance structure in $d \pi^0 \pi^0$ and $d \pi^+ \pi^-$
  systems}},\ }\href {https://doi.org/10.1103/PhysRevC.88.055208} {\bibfield
  {journal} {\bibinfo  {journal} {Phys. Rev. C}\ }\textbf {\bibinfo {volume}
  {88}},\ \bibinfo {pages} {055208} (\bibinfo {year}
  {2013}{\natexlab{b}})}\BibitemShut {NoStop}%
\bibitem [{\citenamefont {P.~Adlarson et al. (WASA-at
  COSY~Collaboration}(2014)}]{d2380}%
  \BibitemOpen
  \bibfield  {author} {\bibinfo {author} {\bibfnamefont {S.~D. A.~C.}\
  \bibnamefont {P.~Adlarson et al. (WASA-at COSY~Collaboration}},\ }\bibfield
  {title} {\bibinfo {title} {{Evidence for a New Resonance from Polarized
  Neutron-Proton Scattering}},\ }\href
  {https://doi.org/10.1103/PhysRevLett.112.202301} {\bibfield  {journal}
  {\bibinfo  {journal} {Phys. Rev. Lett.}\ }\textbf {\bibinfo {volume} {112}},\
  \bibinfo {pages} {202301} (\bibinfo {year} {2014})}\BibitemShut {NoStop}%
\bibitem [{\citenamefont {Xia}\ \emph {et~al.}(2022)\citenamefont {Xia},
  \citenamefont {{\relax S. Fan, X. Zhu and H. Huang}},\ and\ \citenamefont
  {Ping}}]{prc2022}%
  \BibitemOpen
  \bibfield  {author} {\bibinfo {author} {\bibfnamefont {Z.}~\bibnamefont
  {Xia}}, \bibinfo {author} {\bibnamefont {{\relax S. Fan, X. Zhu and H.
  Huang}}},\ and\ \bibinfo {author} {\bibfnamefont {J.}~\bibnamefont {Ping}},\
  }\bibfield  {title} {\bibinfo {title} {{Search for doubly heavy dibaryons in
  the quark delocalization color screening model}},\ }\href
  {https://doi.org/10.1103/PhysRevC.105.025201} {\bibfield  {journal} {\bibinfo
   {journal} {Phys. Rev. C}\ }\textbf {\bibinfo {volume} {105}},\ \bibinfo
  {pages} {025201} (\bibinfo {year} {2022})}\BibitemShut {NoStop}%
\bibitem [{\citenamefont {Huang}\ \emph {et~al.}(2020)\citenamefont {Huang},
  \citenamefont {{\relax J. Ping and X. Zhu}},\ and\ \citenamefont
  {Wang}}]{wang}%
  \BibitemOpen
  \bibfield  {author} {\bibinfo {author} {\bibfnamefont {H.}~\bibnamefont
  {Huang}}, \bibinfo {author} {\bibnamefont {{\relax J. Ping and X. Zhu}}},\
  and\ \bibinfo {author} {\bibfnamefont {F.}~\bibnamefont {Wang}},\ }\bibfield
  {title} {\bibinfo {title} {{Full heavy dibaryons}},\ }\href@noop {} {\
  (\bibinfo {year} {2020})},\ \Eprint {https://arxiv.org/abs/hep-ph/2011.00513}
  {arXiv:hep-ph/2011.00513} \BibitemShut {NoStop}%
\bibitem [{\citenamefont {Y.~Lyu}\ and\ \citenamefont
  {Miyamoto}(2021)}]{prlbaryons}%
  \BibitemOpen
  \bibfield  {author} {\bibinfo {author} {\bibfnamefont {{\relax H. Tong, T.
  Sugiura, S. Aoki, T. Doi, T. Hatsuda and J. Meng}.}~\bibnamefont {Y.~Lyu}}\
  and\ \bibinfo {author} {\bibfnamefont {T.}~\bibnamefont {Miyamoto}},\
  }\bibfield  {title} {\bibinfo {title} {{Dibaryon with Highest Charm Number
  near Unitarity from Lattice QCD}},\ }\href
  {https://doi.org/10.1103/PhysRevLett.127.072003} {\bibfield  {journal}
  {\bibinfo  {journal} {Phys. Rev. Lett.}\ }\textbf {\bibinfo {volume} {127}},\
  \bibinfo {pages} {0072003} (\bibinfo {year} {2021})}\BibitemShut {NoStop}%
\bibitem [{\citenamefont {Wang}(2022)}]{suma}%
  \BibitemOpen
  \bibfield  {author} {\bibinfo {author} {\bibfnamefont {Z.}~\bibnamefont
  {Wang}},\ }\bibfield  {title} {\bibinfo {title} {{Fully-heavy hexaquark
  states via de QCD sum rules}},\ }\href@noop {} {\  (\bibinfo {year}
  {2022})},\ \Eprint {https://arxiv.org/abs/hep-ph/2201.02955}
  {arXiv:hep-ph/2201.02955} \BibitemShut {NoStop}%
\bibitem [{\citenamefont {Richard}\ \emph {et~al.}(2020)\citenamefont
  {Richard}, \citenamefont {{\relax A. Valcarce}},\ and\ \citenamefont
  {Vijande}}]{vijande}%
  \BibitemOpen
  \bibfield  {author} {\bibinfo {author} {\bibfnamefont {J.}~\bibnamefont
  {Richard}}, \bibinfo {author} {\bibnamefont {{\relax A. Valcarce}}},\ and\
  \bibinfo {author} {\bibfnamefont {J.}~\bibnamefont {Vijande}},\ }\bibfield
  {title} {\bibinfo {title} {{Very heavy flavored dibaryons}},\ }\href
  {https://doi.org/10.1103/PhysRevLett.124.212001} {\bibfield  {journal}
  {\bibinfo  {journal} {Phys. Rev. Lett.}\ }\textbf {\bibinfo {volume} {124}},\
  \bibinfo {pages} {212001} (\bibinfo {year} {2020})}\BibitemShut {NoStop}%
\bibitem [{\citenamefont {L\"u}\ \emph {et~al.}(2022)\citenamefont {L\"u},
  \citenamefont {{\relax D. Y. Chen}},\ and\ \citenamefont {Dong}}]{hexa22}%
  \BibitemOpen
  \bibfield  {author} {\bibinfo {author} {\bibfnamefont {Q.}~\bibnamefont
  {L\"u}}, \bibinfo {author} {\bibnamefont {{\relax D. Y. Chen}}},\ and\
  \bibinfo {author} {\bibfnamefont {Y.}~\bibnamefont {Dong}},\ }\bibfield
  {title} {\bibinfo {title} {{Fully-heavy hexaquarks in a constituent quark
  model}},\ }\href@noop {} {\  (\bibinfo {year} {2022})},\ \Eprint
  {https://arxiv.org/abs/hep-ph/2208.03041} {arXiv:hep-ph/2208.03041}
  \BibitemShut {NoStop}%
\bibitem [{\citenamefont {Weng}\ and\ \citenamefont {Zhu}(2022)}]{hexa22b}%
  \BibitemOpen
  \bibfield  {author} {\bibinfo {author} {\bibfnamefont {X.}~\bibnamefont
  {Weng}}\ and\ \bibinfo {author} {\bibfnamefont {S.}~\bibnamefont {Zhu}},\
  }\bibfield  {title} {\bibinfo {title} {{Systematics of fully-heavy
  dibaryons}},\ }\href@noop {} {\  (\bibinfo {year} {2022})},\ \Eprint
  {https://arxiv.org/abs/hep-ph/2207.05505} {arXiv:hep-ph/2207.05505}
  \BibitemShut {NoStop}%
\bibitem [{\citenamefont {et~al (LHCb~collaboration)}(2020)}]{x6900}%
  \BibitemOpen
  \bibfield  {author} {\bibinfo {author} {\bibfnamefont {R.~A.}\ \bibnamefont
  {et~al (LHCb~collaboration)}},\ }\bibfield  {title} {\bibinfo {title}
  {{Observation of structure in the J/$\Psi$-pair mass spectrum}},\ }\href
  {https://doi.org/10.1016/j.scib.2020.08.032} {\bibfield  {journal} {\bibinfo
  {journal} {Sci. Bull.}\ }\textbf {\bibinfo {volume} {65}},\ \bibinfo {pages}
  {1983} (\bibinfo {year} {2020})}\BibitemShut {NoStop}%
\bibitem [{\citenamefont {Semay}\ and\ \citenamefont
  {Silvestre-Brac}(1994)}]{Semay:1994ht}%
  \BibitemOpen
  \bibfield  {author} {\bibinfo {author} {\bibfnamefont {C.}~\bibnamefont
  {Semay}}\ and\ \bibinfo {author} {\bibfnamefont {B.}~\bibnamefont
  {Silvestre-Brac}},\ }\bibfield  {title} {\bibinfo {title} {{Diquonia and
  potential models}},\ }\href {https://doi.org/10.1007/BF01413104} {\bibfield
  {journal} {\bibinfo  {journal} {Z. Phys. C}\ }\textbf {\bibinfo {volume}
  {61}},\ \bibinfo {pages} {271} (\bibinfo {year} {1994})}\BibitemShut
  {NoStop}%
\bibitem [{\citenamefont {Silvestre-Brac}(1996)}]{SilvestreBrac:1996bg}%
  \BibitemOpen
  \bibfield  {author} {\bibinfo {author} {\bibfnamefont {B.}~\bibnamefont
  {Silvestre-Brac}},\ }\bibfield  {title} {\bibinfo {title} {{Spectrum and
  static properties of heavy baryons}},\ }\href
  {https://doi.org/10.1007/s006010050028} {\bibfield  {journal} {\bibinfo
  {journal} {Few Body Syst.}\ }\textbf {\bibinfo {volume} {20}},\ \bibinfo
  {pages} {1} (\bibinfo {year} {1996})}\BibitemShut {NoStop}%
\bibitem [{\citenamefont {Gordillo}\ \emph {et~al.}(2020)\citenamefont
  {Gordillo}, \citenamefont {{\relax F De Soto}},\ and\ \citenamefont
  {Segovia}}]{prdyo1}%
  \BibitemOpen
  \bibfield  {author} {\bibinfo {author} {\bibfnamefont {M.~C.}\ \bibnamefont
  {Gordillo}}, \bibinfo {author} {\bibnamefont {{\relax F De Soto}}},\ and\
  \bibinfo {author} {\bibfnamefont {J.}~\bibnamefont {Segovia}},\ }\bibfield
  {title} {\bibinfo {title} {{Diffusion Monte Carlo calculations of fully-heavy
  multiquark bound states}},\ }\href
  {https://doi.org/10.1103/PhysRevD.102.114007} {\bibfield  {journal} {\bibinfo
   {journal} {Phys. Rev. D}\ }\textbf {\bibinfo {volume} {102}},\ \bibinfo
  {pages} {111007} (\bibinfo {year} {2020})}\BibitemShut {NoStop}%
\bibitem [{\citenamefont {Kalos}(1970)}]{kalos}%
  \BibitemOpen
  \bibfield  {author} {\bibinfo {author} {\bibfnamefont {M.~H.}\ \bibnamefont
  {Kalos}},\ }\bibfield  {title} {\bibinfo {title} {{Energy of a Boson Fluid
  with Lennard-Jones Potentials}},\ }\href
  {https://doi.org/10.1103/PhysRevA.2.250} {\bibfield  {journal} {\bibinfo
  {journal} {Phys. Rev. A}\ }\textbf {\bibinfo {volume} {2}},\ \bibinfo {pages}
  {250} (\bibinfo {year} {1970})}\BibitemShut {NoStop}%
\bibitem [{\citenamefont {Boronat}\ and\ \citenamefont
  {Casulleras}(1994)}]{boro94}%
  \BibitemOpen
  \bibfield  {author} {\bibinfo {author} {\bibfnamefont {J.}~\bibnamefont
  {Boronat}}\ and\ \bibinfo {author} {\bibfnamefont {J.}~\bibnamefont
  {Casulleras}},\ }\bibfield  {title} {\bibinfo {title} {{Monte Carlo analysis
  of an interatomic potential for He}},\ }\href
  {https://doi.org/10.1103/PhysRevB.49.8920} {\bibfield  {journal} {\bibinfo
  {journal} {Phys. Rev. B}\ }\textbf {\bibinfo {volume} {49}},\ \bibinfo
  {pages} {8920} (\bibinfo {year} {1994})}\BibitemShut {NoStop}%
\bibitem [{\citenamefont {Hammond}\ \emph {et~al.}(1994)\citenamefont
  {Hammond}, \citenamefont {Lester},\ and\ \citenamefont {Reynolds}}]{hammond}%
  \BibitemOpen
  \bibfield  {author} {\bibinfo {author} {\bibfnamefont {B.}~\bibnamefont
  {Hammond}}, \bibinfo {author} {\bibfnamefont {W.}~\bibnamefont {Lester}},\
  and\ \bibinfo {author} {\bibfnamefont {P.}~\bibnamefont {Reynolds}},\
  }\href@noop {} {\emph {\bibinfo {title} {Monte Carlo Methods in ab Initio
  Quantum Chemistry}}}\ (\bibinfo  {publisher} {World Scientific},\ \bibinfo
  {address} {Singapore},\ \bibinfo {year} {1994})\BibitemShut {NoStop}%
\bibitem [{\citenamefont {S\'anchez-Baena}\ \emph {et~al.}(2018)\citenamefont
  {S\'anchez-Baena}, \citenamefont {Boronat},\ and\ \citenamefont
  {Mazzanti}}]{spin-orbita}%
  \BibitemOpen
  \bibfield  {author} {\bibinfo {author} {\bibfnamefont {J.}~\bibnamefont
  {S\'anchez-Baena}}, \bibinfo {author} {\bibfnamefont {J.}~\bibnamefont
  {Boronat}},\ and\ \bibinfo {author} {\bibfnamefont {F.}~\bibnamefont
  {Mazzanti}},\ }\bibfield  {title} {\bibinfo {title} {{Diffusion Monte Carlo
  methods for spin-orbit-coupled ultracold Bose gases}},\ }\href
  {https://doi.org/10.1103/PhysRevA.98.053632} {\bibfield  {journal} {\bibinfo
  {journal} {Phys. Rev. A}\ }\textbf {\bibinfo {volume} {98}},\ \bibinfo
  {pages} {053632} (\bibinfo {year} {2018})}\BibitemShut {NoStop}%
\bibitem [{\citenamefont {Goldman}\ \emph {et~al.}(1989)\citenamefont
  {Goldman}, \citenamefont {{\relax K. Maltman, G.E. Stephenson, Jr., K.E.
  Schmidt}},\ and\ \citenamefont {Wang}}]{fivefunctions}%
  \BibitemOpen
  \bibfield  {author} {\bibinfo {author} {\bibfnamefont {T.}~\bibnamefont
  {Goldman}}, \bibinfo {author} {\bibnamefont {{\relax K. Maltman, G.E.
  Stephenson, Jr., K.E. Schmidt}}},\ and\ \bibinfo {author} {\bibfnamefont
  {F.}~\bibnamefont {Wang}},\ }\bibfield  {title} {\bibinfo {title}
  {{"Inevitable" nonstrange dibaryon}},\ }\href@noop {} {\bibfield  {journal}
  {\bibinfo  {journal} {Phys. Rev. C}\ }\textbf {\bibinfo {volume} {39}},\
  \bibinfo {pages} {1889} (\bibinfo {year} {1989})}\BibitemShut {NoStop}%
\bibitem [{\citenamefont {Brodsky}\ \emph {et~al.}(1983)\citenamefont
  {Brodsky}, \citenamefont {{\relax C.R. Ji}},\ and\ \citenamefont
  {Wang}}]{5f2}%
  \BibitemOpen
  \bibfield  {author} {\bibinfo {author} {\bibfnamefont {S.}~\bibnamefont
  {Brodsky}}, \bibinfo {author} {\bibnamefont {{\relax C.R. Ji}}},\ and\
  \bibinfo {author} {\bibfnamefont {F.}~\bibnamefont {Wang}},\ }\bibfield
  {title} {\bibinfo {title} {{Quantum Chromodynamic predictions for the
  deuteron form factor }},\ }\href@noop {} {\bibfield  {journal} {\bibinfo
  {journal} {Phys. Rev. Lett.}\ }\textbf {\bibinfo {volume} {51}},\ \bibinfo
  {pages} {83} (\bibinfo {year} {1983})}\BibitemShut {NoStop}%
\bibitem [{\citenamefont {Ji}\ and\ \citenamefont {Brodsky}(1986)}]{5f3}%
  \BibitemOpen
  \bibfield  {author} {\bibinfo {author} {\bibfnamefont {C.-R.}\ \bibnamefont
  {Ji}}\ and\ \bibinfo {author} {\bibfnamefont {S.}~\bibnamefont {Brodsky}},\
  }\bibfield  {title} {\bibinfo {title} {{Quantum-chromodynamic evolution of
  six-quark states}},\ }\href@noop {} {\bibfield  {journal} {\bibinfo
  {journal} {Phys. Rev. D}\ }\textbf {\bibinfo {volume} {34}},\ \bibinfo
  {pages} {1460} (\bibinfo {year} {1986})}\BibitemShut {NoStop}%
\bibitem [{\citenamefont {Kima}\ \emph {et~al.}(2020)\citenamefont {Kima},
  \citenamefont {{\relax K.S. Kim}},\ and\ \citenamefont {Oka}}]{5f0}%
  \BibitemOpen
  \bibfield  {author} {\bibinfo {author} {\bibfnamefont {H.}~\bibnamefont
  {Kima}}, \bibinfo {author} {\bibnamefont {{\relax K.S. Kim}}},\ and\ \bibinfo
  {author} {\bibfnamefont {M.}~\bibnamefont {Oka}},\ }\bibfield  {title}
  {\bibinfo {title} {{Hexaquark picture for d$^*$(2380)}},\ }\href
  {https://doi.org/10.1103/PhysRevD.102.074023} {\bibfield  {journal} {\bibinfo
   {journal} {Phys. Rev. D}\ }\textbf {\bibinfo {volume} {102}},\ \bibinfo
  {pages} {074023} (\bibinfo {year} {2020})}\BibitemShut {NoStop}%
\bibitem [{\citenamefont {Carlson}\ \emph {et~al.}(2015)\citenamefont
  {Carlson}, \citenamefont {{\relax S. Gandolfi, F. Pederiva, S.C. Pieper, R.
  Schiavilla, K.E. Schmidt}},\ and\ \citenamefont {Wiringa}}]{carlson}%
  \BibitemOpen
  \bibfield  {author} {\bibinfo {author} {\bibfnamefont {J.}~\bibnamefont
  {Carlson}}, \bibinfo {author} {\bibnamefont {{\relax S. Gandolfi, F.
  Pederiva, S.C. Pieper, R. Schiavilla, K.E. Schmidt}}},\ and\ \bibinfo
  {author} {\bibfnamefont {R.}~\bibnamefont {Wiringa}},\ }\bibfield  {title}
  {\bibinfo {title} {{Quantum Monte Carlo methods for nuclear physics}},\
  }\href {https://doi.org/101103/RevModPhys.87.1067} {\bibfield  {journal}
  {\bibinfo  {journal} {Rev. Mod. Phys}\ }\textbf {\bibinfo {volume} {87}},\
  \bibinfo {pages} {1067} (\bibinfo {year} {2015})}\BibitemShut {NoStop}%
\bibitem [{\citenamefont {Gordillo}\ \emph {et~al.}(2021)\citenamefont
  {Gordillo}, \citenamefont {{\relax F. De Soto}},\ and\ \citenamefont
  {Segovia}}]{prdyo2}%
  \BibitemOpen
  \bibfield  {author} {\bibinfo {author} {\bibfnamefont {M.~C.}\ \bibnamefont
  {Gordillo}}, \bibinfo {author} {\bibnamefont {{\relax F. De Soto}}},\ and\
  \bibinfo {author} {\bibfnamefont {J.}~\bibnamefont {Segovia}},\ }\bibfield
  {title} {\bibinfo {title} {{Structure of the X(3872) as explained by a
  diffusion Monte Carlo calculation}},\ }\href
  {https://doi.org/10.1103/PhysRevD.104.054036} {\bibfield  {journal} {\bibinfo
   {journal} {Phys. Rev. D}\ }\textbf {\bibinfo {volume} {104}},\ \bibinfo
  {pages} {054036} (\bibinfo {year} {2021})}\BibitemShut {NoStop}%
\bibitem [{\citenamefont {Hiyama}\ \emph {et~al.}(2003)\citenamefont {Hiyama},
  \citenamefont {Kino},\ and\ \citenamefont {Kamikura}}]{hiyama}%
  \BibitemOpen
  \bibfield  {author} {\bibinfo {author} {\bibfnamefont {E.}~\bibnamefont
  {Hiyama}}, \bibinfo {author} {\bibfnamefont {Y.}~\bibnamefont {Kino}},\ and\
  \bibinfo {author} {\bibfnamefont {M.}~\bibnamefont {Kamikura}},\ }\bibfield
  {title} {\bibinfo {title} {{Gaussian expansion method for few-body
  systems}},\ }\href@noop {} {\bibfield  {journal} {\bibinfo  {journal} {Prog.
  Part. Nucl. Phys.}\ }\textbf {\bibinfo {volume} {51}},\ \bibinfo {pages}
  {223} (\bibinfo {year} {2003})}\BibitemShut {NoStop}%
\end{thebibliography}%

\end{document}